\documentclass[lettersize,journal]{IEEEtran}
\usepackage{amsmath,amsfonts,amssymb}
\usepackage{algpseudocode}
\usepackage{algorithm}
\usepackage{array}
\usepackage[caption=false,font=footnotesize,labelfont=rm,textfont=rm]{subfig}
\usepackage{bbding}
\usepackage{mathrsfs}
\usepackage{wasysym}
\usepackage{textcomp}
\usepackage{stfloats}
\newtheorem{proposition}{Proposition}
\usepackage{url}
\usepackage{verbatim}
\usepackage{graphicx}
\usepackage{cite}
\usepackage{multirow}
\usepackage{booktabs}
\usepackage{xcolor}
\usepackage{makecell}
\usepackage[T1]{fontenc}
\def\BibTeX{{\rm B\kern-.05em{\sc i\kern-.025em b}\kern-.08em
    T\kern-.1667em\lower.7ex\hbox{E}\kern-.125emX}}

\begin{document}

\title{Starlight: Whittle-Driven Fidelity Rescue for Satellite-Terrestrial Quantum Networks}
%\noindent\textbf{Naming.} We call our protocol \emph{Starlight} to reflect that satellite-assisted entanglement distribution is ultimately carried by photons and leverages space-borne links to ``rescue'' degraded inter-domain bottlenecks.
%Starlight turns LEO satellites from supply-driven entanglement providers into demand-driven fidelity rescuers, using EAoS to quantify physical degradation and Whittle-style scheduling to prioritize the most urgent inter-domain bottlenecks in real time.

\author{Yaliang Shi,~Zhiwei Zhao,~Bangguo Yuan
        % <-this % stops a space
\thanks{Yaliang Shi, Zhiwei Zhao, and Bangguo Yuan are with the School of Computer Science and Engineering, University of Electronic Science and Technology of China, Chengdu, Sichuan 611731, China (e-mail: \{yaliang,zhiwei,bangguo\}@mobinets.org).}}



\maketitle

\begin{abstract}
A planetary-scale quantum internet is constrained by exponential attenuation in terrestrial fibers. Low-Earth-orbit (LEO) satellites provide a low-loss vacuum backbone, but limited satellite resources must be allocated among competing connections under continuously evolving and heterogeneous physical conditions. Existing satellite scheduling remains largely provisioning-oriented, optimizing satellite-resource assignment for aggregate entanglement delivery. Recent age-based approaches instead incorporate chronological freshness into entanglement scheduling. Neither line of work directly identifies the connections most urgently requiring satellite assistance as their physical states evolve. We propose Starlight, a demand-driven satellite--terrestrial entanglement routing architecture that targets degrading connections before their terrestrial entanglement becomes unusable. Starlight introduces Effective Age of State (EAoS), a physics-aware metric that maps heterogeneous decoherence and swapping-induced fidelity penalties onto a unified additive coordinate. Representing each connection by evolving EAoS, we formulate rescue scheduling as a Restless Multi-Armed Bandit (RMAB) problem and derive a closed-form Whittle-style index approximation for top-$K$ selection. Starlight further employs a two-timescale control plane and an asynchronous dispatch-and-mask pipeline to suppress redundant in-flight allocations. Simulations show that Starlight achieves up to $84.5\%$ average satisfaction under constrained rescue capacity and sub-$0.5$-ms orchestration at $1{,}000$ connections, while remaining robust to heterogeneous memory decay and stochastic atmospheric degradation.
\end{abstract}

\begin{IEEEkeywords}
Quantum Internet, Satellite-Terrestrial Networks, Entanglement Routing, Restless Multi-Armed Bandit (RMAB), Fidelity-Aware Scheduling
\end{IEEEkeywords}

\section{Introduction}
\label{sec:intro}
\IEEEPARstart{T}{he} distribution of high-fidelity end-to-end (E2E) entanglement is the bedrock of the global quantum internet \cite{wehner2018quantum} and is required to faithfully teleport quantum states between distant nodes for transformative applications such as distributed quantum computing, quantum sensing, and quantum-secure communications \cite{khalid2024integrated}. Scaling this vision across purely terrestrial deployments confronts a severe physical limit: photon loss in optical fibers scales exponentially with distance ($\sim 0.2$ dB/km) \cite{yin2017satellite}. Although terrestrial quantum repeaters use entanglement swapping to extend entanglement across multiple hops and purification to recover degraded fidelity, accumulated operational errors and purification overheads force these multi-hop chains into a practical ``Fidelity Wall'' over long-haul transmissions \cite{goswami2025satellites}. Consequently, terrestrial routing architectures remain difficult to scale to planetary dimensions without severe degradation in distribution rates.

\begin{figure*}[t] 
    \centering
    \includegraphics[width=\linewidth]{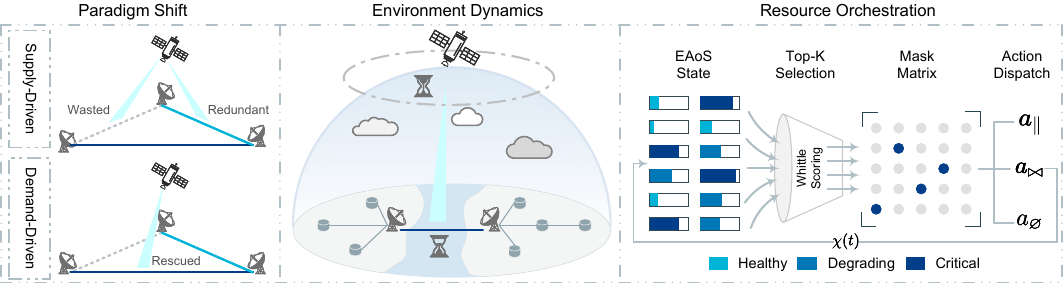} 
    \caption{\textbf{Starlight cooperative architecture.} \textbf{(a)} Demand-driven targeting mitigates the resource mismatch of static supply-driven paradigms. Cyan beams denote satellite downlink attempts carrying entangled photons, while gray dashed links denote expired entanglement. Cyan, blue, and dark blue indicate healthy, degrading, and critical EAoS states, respectively. \textbf{(b)} Dual physical constraints: stochastic atmospheric fading and heterogeneous memory decoherence. \textbf{(c)} The Whittle-Driven Orchestration (WDO) closed-loop control plane computes Whittle-style rescue indices from EAoS states and action feasibility, applies masking to avoid redundant in-flight allocations, and dispatches actions ($a_\varnothing, a_{\parallel}, a_{\bowtie}$) using asynchronous binary feedback $\chi$.}
    \label{fig:architecture}
\end{figure*}

Leveraging near-zero vacuum attenuation, low-Earth-orbit (LEO) satellite constellations bypass terrestrial limits to establish a global low-loss backbone \cite{de2023satellite, ren2017ground, liao2018satellite}. Yet, orchestrating such a hybrid network requires control decisions to keep pace with rapidly evolving quantum states \cite{jiang2025quantum, shi2026efficient}. Existing work predominantly treats space-borne resource allocation as an entanglement-provisioning problem, optimizing satellite-resource assignment for aggregate entanglement delivery. Centralized offline planners---including ILP-based satellite scheduling \cite{panigrahy2022optimal,wei2024optimal}, time-discretized logical-graph optimization \cite{chang2024entanglement}, and hybrid quantum-classical solvers \cite{wei2024entanglement}---incur increasing computational cost as the network scales. Moreover, their aggregate-delivery objective does not explicitly capture the evolving rescue urgency of terrestrial connections and may allocate assistance to serviceable but less critical connections, as illustrated in Fig.~\ref{fig:architecture}(a). Planning-based optimization can also create an algorithmic--physical latency gap: terrestrial entanglement states may continue to degrade during planning and execution, rendering precomputed allocations stale or even wasted. In our setting, terrestrial protocols may establish entangled subpaths on both sides of a long-distance inter-domain span while the span itself remains unresolved. We call this ready-but-unbridged state an \emph{inter-domain bottleneck}, and satellite assistance that bridges it before the terrestrial subpaths become unusable \emph{fidelity rescue}.

A natural candidate for representing such dynamic urgency is temporal freshness, as explored by recent quantum-network scheduling metrics such as Age of Entanglement (AoE) \cite{ceran2026age} and Fidelity-Age (FA) \cite{ercetin2026fidelity}. However, chronological age alone is insufficient for state-aware fidelity rescue. Although these approaches account for quantum errors such as decoherence and swapping in their physical models, their scheduling states still evolve primarily with elapsed time and therefore do not directly encode the accumulated physical degradation. The same elapsed time can induce different fidelity losses across heterogeneous terrestrial memories and between terrestrial and space-borne storage \cite{gundougan2021proposal,panahi2025energy}, as illustrated in Fig.~\ref{fig:architecture}(b), while entanglement swapping can cause discrete fidelity drops without a corresponding increase in elapsed time. More fundamentally, existing age-based metrics characterize the freshness of established entanglement, whereas an unresolved inter-domain bottleneck does not yet constitute an established end-to-end entanglement and therefore has no intrinsic entanglement age; its rescue urgency must instead be inferred from the degrading terrestrial subpaths on its two sides. Temporal freshness alone is therefore insufficient to determine when and where satellite intervention is most urgently needed.

To address these limitations, we propose Starlight, a demand-driven satellite--terrestrial entanglement routing architecture that coordinates fidelity rescue across multiple inter-domain bottlenecks. Starlight adopts a decoupled intra/inter-domain design, treating these bottlenecks as dynamic rescue targets. Satellites act as on-demand ``fidelity healers,'' providing assistance when and where terrestrial connections most urgently require rescue. To quantify this urgency, we introduce the Effective Age of State (EAoS), a physics-aware state metric that directly represents accumulated physical degradation instead of using elapsed time as the scheduling state. Through a logarithmic projection, EAoS maps heterogeneous decoherence and swapping-induced fidelity penalties onto a unified additive scale, enabling direct comparison across heterogeneous bottleneck states.

With each inter-domain bottleneck represented by an evolving EAoS state and competing for shared satellite rescue capacity, we formulate global rescue admission as a Restless Multi-Armed Bandit (RMAB) problem. Since general RMABs are PSPACE-hard~\cite{guha2010approximation}, directly solving the coupled problem becomes computationally prohibitive as the network scales. We therefore apply a Whittle-style Lagrangian relaxation to decompose it into independent per-bottleneck subproblems and derive a closed-form rescue index approximation for scalable prioritization. Starlight then adopts a two-timescale control architecture: terrestrial Anchors report EAoS and atmospheric telemetry, while the orchestrator performs top-$K$ selection for real-time rescue triage, as illustrated in Fig.~\ref{fig:architecture}(c). An asynchronous dispatch-and-mask pipeline further decouples real-time orchestration from variable-latency physical execution, keeping in-flight bottlenecks masked during operation and heralding delays to prevent redundant allocations.

In summary, our main contributions are as follows:
\begin{itemize}
    \item We formulate satellite-assisted entanglement routing from a demand-driven fidelity-rescue perspective, prioritizing scarce satellite assistance according to the evolving rescue urgency of dynamic inter-domain bottlenecks.
    
    \item We introduce the Effective Age of State (EAoS), a physics-aware degradation state that unifies heterogeneous decoherence and swapping-induced fidelity penalties on an additive scale.
    
    \item We formulate capacity-constrained rescue admission as a Restless Multi-Armed Bandit (RMAB) problem and derive a closed-form Whittle-style index approximation for scalable top-$K$ rescue selection.
    
    \item We develop a two-timescale control plane with an asynchronous dispatch-and-mask pipeline for variable-latency physical execution and evaluate Starlight under resource scarcity, memory heterogeneity, stochastic atmospheric degradation, and network scaling, with exact value-iteration comparison and component ablations.
\end{itemize}

\section{Related Work}
\textbf{Terrestrial Routing and Physical Limits.} Terrestrial quantum routing has evolved from centralized SDN-style control and network-wide resource orchestration \cite{shi2020concurrent, zhao2021redundant, yang2023asynchronous} toward P4-based programmable frameworks such as QuIP \cite{kozlowski2024quip} and decentralized routing or consensus mechanisms such as Q-DDCA \cite{chen2023q} and REDP \cite{chen2024redp}. Despite these advances, exponential fiber attenuation ($\sim 0.2$~dB/km) \cite{yin2017satellite} still imposes a long-haul ``Fidelity Wall'' \cite{goswami2025satellites}, motivating low-Earth-orbit (LEO) satellites as active inter-domain bridges.

\textbf{Space Architectures and Resource Provisioning.} To bypass terrestrial limits, prior work has explored integrated satellite--terrestrial quantum networks \cite{conti2024satellite}, mobile control planes \cite{chiti2022mobile}, and satellite-assisted multi-domain architectures \cite{huang2023socially}. Space-borne resource allocation was initially formulated through integer-programming models \cite{panigrahy2022optimal, wei2024optimal} to maximize aggregate entanglement distribution, while subsequent studies incorporated LEO mobility through temporal discretization. Chang \textit{et al.} \cite{chang2024entanglement} transformed time-varying topologies into static logical graphs, whereas QuESat \cite{gu2025quesat} provisioned lightpaths through randomized and deterministic rounding. The growing combinatorial complexity of such formulations has further motivated scalable greedy heuristics \cite{williams2024scalable} and hybrid quantum-classical solvers based on Dantzig-Wolfe decomposition \cite{wei2024entanglement}. Despite this progression, these methods remain primarily provisioning-oriented, optimizing aggregate entanglement delivery rather than explicitly prioritizing degrading terrestrial connections by rescue urgency.

\textbf{Age-Aware Metrics and Rescue Urgency.} Recent quantum network scheduling has explored time-driven freshness metrics, including Age of Entanglement (AoE) \cite{ceran2026age} and Fidelity-Age (FA) scheduling \cite{ercetin2026fidelity}. AoE derives dynamic policies through relative value iteration over a repeater-chain MDP, whereas FA-aware scheduling improves tractability through lightweight threshold and Lyapunov-drift-inspired scoring. Despite these advances, their scheduling states remain primarily tied to chronological freshness and do not directly encode heterogeneous memory decay and operation-induced fidelity loss within a unified physical degradation state. More fundamentally, these formulations characterize entanglement freshness rather than the evolving rescue urgency of unresolved inter-domain bottlenecks. In particular, AoE optimizes continuous entanglement-distribution freshness within an individual repeater chain, whereas Starlight coordinates physics-aware rescue across multiple evolving bottlenecks competing for shared satellite capacity.

\section{System Model and Problem Formulation}
\subsection{Network Architecture and Physical Constraints}
\label{subsec:architecture}

We model the hybrid network as a dynamic graph $\mathcal{G}\triangleq(\mathcal{V},\mathcal{E})$, where terrestrial vertices $\mathcal{V}_{\mathrm{G}}\subset\mathcal{V}$ comprise intra-domain repeaters and Anchor Nodes, satellite vertices $\mathcal{V}_{\mathrm{S}}\subset\mathcal{V}$ represent the LEO constellation, and $\mathcal{E}=\mathcal{E}_{\mathrm{G2G}}\cup\mathcal{E}_{\mathrm{S2G}}$ contains terrestrial fibers and space-to-ground downlinks\footnote{We restrict $\mathcal{E}_{\mathrm{S2G}}$ to downlinks, as terrestrial-to-space quantum uplinks suffer from prohibitive beam broadening and diffraction losses induced by near-source atmospheric turbulence \cite{ren2017ground}.}.

Starlight adopts a decoupled intra/inter-domain architecture in which each terrestrial domain is a locally routable repeater subnetwork managed by an underlying entanglement-establishment protocol. We index persistent inter-domain bottleneck flows by $\mathcal{M}$. For each bottleneck $m\in\mathcal{M}$, the terrestrial protocols expose successive ready-but-unbridged rescue instances consisting of one established intra-domain entangled subpath on each side of an unresolved inter-Anchor span. Starlight bridges this span before the subpaths become unusable; successful completion or physical collapse terminates the current instance, after which the next instance is exposed.

Each LEO satellite $v\in\mathcal{V}_{\mathrm{S}}$ has $T_v$ transmitters and $Q_v$ quantum memories, while each Anchor Node $u\in\mathcal{V}_{\mathrm{G}}$ has $R_u$ optical receivers, consistent with prior satellite-entanglement resource models \cite{jennewein2025qeyssat,panigrahy2022optimal,wei2024entanglement}. Anchor-side quantum-memory capacity is managed by the underlying terrestrial domains and is therefore excluded from the rescue-layer resource constraints. Beyond these hardware limits, space-borne rescue exhibits heterogeneous physical characteristics: terrestrial and space-borne quantum memories may exhibit different decoherence rates, denoted by $\gamma_{\mathrm{G}}$ and $\gamma_{\mathrm{S}}$, respectively, while terrestrial and in-orbit Bell State Measurements (BSMs) for entanglement swapping incur fidelity penalties $P_{\mathrm{G}}$ and $P_{\mathrm{S}}$, respectively \cite{belenchia2022quantum}. Together with atmospheric loss over $\mathcal{E}_{\mathrm{S2G}}$, these physical factors motivate the heterogeneous rescue modes defined in Section~\ref{subsec:action_space}.

\subsection{State Space $\mathcal{S}$: The Effective Age of State (EAoS)}

To represent the physical urgency of unresolved inter-domain rescue, we define the Effective Age of State (EAoS) as an additive degradation coordinate. For an entanglement segment $c$ with fidelity $F_c(t)$, let $\lambda_c(t)\triangleq(4F_c(t)-1)/3$ denote its depolarizing parameter under the Werner abstraction \cite{werner1989quantum}. This representation is convenient because multiplicative degradation of the entanglement quality under storage and swapping can be converted into additive degradation through a logarithmic projection. Relative to the source parameter $\lambda_0$, its EAoS is $e_c(t)\triangleq\ln(\lambda_0/\lambda_c(t))$. A larger EAoS therefore corresponds to greater accumulated physical degradation and lower entanglement quality. Let $t_0$ denote the epoch at which Starlight begins tracking segment $c$, and let $\Psi_c(t_0,t]$ denote the degradative quantum operations modeled thereafter. For operation $k$, define $P_k\triangleq-\ln\eta_k\geq0$, where $\eta_k\in(0,1]$ is its multiplicative shrinkage factor. Then
\begin{equation} \label{eq:eaos_definition}
    e_c(t)
    =
    e_c(t_0)
    +
    \int_{t_0}^{t} \gamma_c(\tau) d\tau
    +
    \sum_{k \in \Psi_c(t_0,t]} P_k .
\end{equation}

The logarithmic projection converts continuous memory decoherence and discrete logic-induced fidelity loss into additive terms, where $\gamma_c(\tau)\in\{\gamma_{\mathrm{G}},\gamma_{\mathrm{S}}\}$ denotes the environment-dependent decay rate and $\gamma_{\mathrm{G}}$ becomes bottleneck-specific when heterogeneous terrestrial memories are considered. For example, an in-orbit BSM incurs $P_k=P_{\mathrm{S}}$. This abstraction therefore tracks the physical degradation observed by Starlight without reconstructing the internal preparation history of the terrestrial domains.

\begin{figure}[!t]
    \centering
    \includegraphics[width=\linewidth]{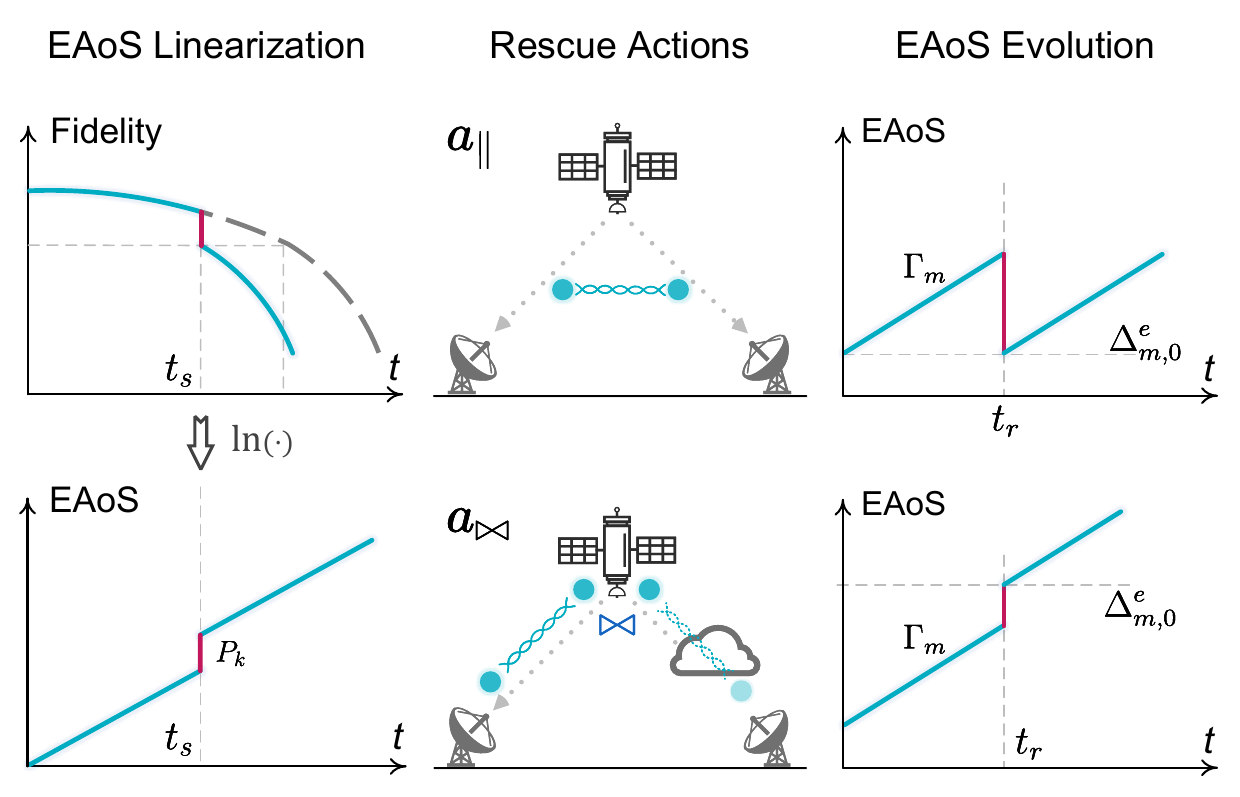}
    \caption{\textbf{EAoS model.} \textbf{(a)} Fidelity linearization to EAoS with swap penalty $P_k$ at $t_s$. \textbf{(b)} Rescue modes $a_{\parallel}$ and $a_{\bowtie}$ for establishing the inter-Anchor entangled bridge. \textbf{(c)} Bottleneck EAoS evolution with drift $\Gamma_m$ and state renewal to $\Delta_{m,0}^e$ at renewal epoch $t_r$.}
    \label{fig:eaos}
\end{figure}

For bottleneck $m\in\mathcal{M}$, let $c_m^{u}$ and $c_m^{w}$ denote the two ready intra-domain subpaths terminating at Anchor Nodes $u_m$ and $w_m$, respectively. Their states at exposure already reflect any native routing, swapping, and purification performed by the underlying terrestrial protocols. Let $\Delta_{\mathrm{req}}^e$ denote the application-level usability threshold and $\Delta_{\mathrm{crit}}^e>\Delta_{\mathrm{req}}^e$ the collapse threshold. We define the bottleneck EAoS as
\begin{equation} \label{eq:macroscopic_eaos}
    \Delta_m^e(t)
    \triangleq
    \min\left\{
    e_{c_m^{u}}(t)+e_{c_m^{w}}(t),
    \Delta_{\mathrm{crit}}^e
    \right\}.
\end{equation}

Anchor Nodes report these scalar states through classical telemetry, allowing the orchestrator to maintain $\mathbf{S}(t)=[\Delta_1^e(t),\dots,\Delta_{|\mathcal{M}|}^e(t)]$ without exposing intra-domain topology or preparation history.

\subsection{Action Space $\mathcal{A}$: Heterogeneous Space-Borne Interventions}
\label{subsec:action_space}

Using the global state $\mathbf{S}(t)$, the orchestrator selects among three space-borne interventions for bottleneck $m$, with each active action establishing an entangled bridge between Anchor Nodes $u_m$ and $w_m$ that is subsequently connected to the two ready intra-domain subpaths by terrestrial swapping. We restrict each active meta-action to a single-satellite realization and define
\[
\mathcal{A}\triangleq\{a_\varnothing,a_{\parallel},a_{\bowtie}\},
\]
as illustrated in Fig.~\ref{fig:eaos}(b).

\textbf{Synchronous Downlink ($a_{\parallel}$):} A satellite simultaneously visible to both Anchors directly distributes the two halves of an entangled pair to $u_m$ and $w_m$. This mode requires neither in-orbit BSM nor space-memory waiting, but depends on simultaneous dual-Anchor coverage and both atmospheric downlinks.

\textbf{In-Orbit Stitching ($a_{\bowtie}$):} A satellite establishes the two satellite--Anchor entanglements across separate visibility opportunities, buffers the first satellite-side state, and performs one in-orbit BSM after the second link becomes available. This mode relaxes simultaneous coverage at the cost of additional synchronization decoherence and the space-borne logic penalty $P_{\mathrm{S}}$.

\textbf{Passive Idling ($a_\varnothing$):} No rescue is attempted in the current micro-slot, and the bottleneck continues its passive EAoS evolution.

\subsection{Transition Dynamics and Rescue Modes}
\label{subsec:transition_dynamics}

We model network evolution over discrete micro-slots of duration $\delta t$. While a ready-but-unbridged instance remains unresolved, its two terrestrial subpaths continue to decohere. Let $\Gamma_m\triangleq\gamma_m^u+\gamma_m^w$ denote the aggregate passive drift rate. The one-step passive drift is
\begin{equation}
    \phi_m(x)
    \triangleq
    \min\left\{
    x+\Gamma_m\delta t,
    \Delta_{\mathrm{crit}}^e
    \right\}.
\end{equation}
Accordingly, let $\Delta_m^{e+}(t)\triangleq\phi_m(\Delta_m^e(t))$ denote the passive successor state.

For each active rescue mode $a$, let $\psi_{m,a}$ denote the prospective EAoS penalty required for fidelity-qualified E2E completion. Synchronous downlink incurs only the two terrestrial Anchor BSM penalties, whereas in-orbit stitching additionally accounts for synchronization-induced decoherence and one in-orbit BSM penalty $P_{\mathrm{S}}$:
\begin{equation} \label{eq:action_penalty}
    \psi_{m,a}
    \triangleq
    \begin{cases}
        2P_{\mathrm{G}},
        & a=a_{\parallel},\\[2pt]
        2P_{\mathrm{G}}+P_{\mathrm{S}}+\left(\Gamma_m+\gamma_{\mathrm{G}}+\gamma_{\mathrm{S}}\right)\bar{\tau}_m,
        & a=a_{\bowtie},
    \end{cases}
\end{equation}
where $\bar{\tau}_m$ denotes the expected asynchronous synchronization delay. The penalty $\psi_{m,a}$ is used for dispatch-time fidelity feasibility and is not separately accumulated during physical execution, avoiding double counting of in-flight degradation.

The fidelity-feasible active-action set is
\begin{equation} \label{eq:action_feasibility}
    \mathcal{A}_m^+(t)
    \triangleq
    \left\{
    a\in\{a_{\parallel},a_{\bowtie}\}:
    \Delta_m^e(t)+\psi_{m,a}\leq\Delta_{\mathrm{req}}^e
    \right\}.
\end{equation}
Geometry and atmospheric conditions are captured separately by the successful-completion probability $p_a(m,t)$, with $p_a(m,t)=0$ when no physical realization is available. If $\mathcal{A}_m^+(t)=\emptyset$, the instance receives no active rescue and continues its passive evolution.

For a pre-collapse state $\Delta_m^e(t)<\Delta_{\mathrm{crit}}^e$ and an admitted action $a_m(t)\in\mathcal{A}_m^+(t)$, let $U_m(t)\sim\mathrm{Bernoulli}(p_{a_m(t)}(m,t))$ denote the completion outcome. A successful rescue renews the persistent bottleneck to an action-independent state $\Delta_{m,0}^e\sim\mathcal{D}_m$, whereas failure leaves the current instance on its passive trajectory. Hence,
\begin{equation} \label{eq:transition_dynamics}
\Delta_m^e(t+1)
=
\Delta_m^{e+}(t)
+
U_m(t)\bigl(
\Delta_{m,0}^e-\Delta_m^{e+}(t)
\bigr).
\end{equation}
At the terminal collapse state $\Delta_m^e(t)=\Delta_{\mathrm{crit}}^e$, the current instance terminates and renews according to
\begin{equation} \label{eq:collapse_renewal}
\Delta_m^e(t+1)
=
\Delta_{m,0}^e.
\end{equation}
For passive idling $a_m(t)=a_\varnothing$, we set $U_m(t)=0$ in Eq.~(\ref{eq:transition_dynamics}).

These transitions define the completion-embedded admission abstraction used by the RMAB. Multi-slot physical execution is handled separately by the asynchronous pipeline in Section~\ref{sec:implementation}, which masks in-flight instances and invalidates pending feedback if the current instance collapses before completion.

\subsection{Problem Formulation: The Global RMAB Paradigm}
\label{subsec:rmab_formulation}

We formulate admission-level orchestration as a discrete-time Restless Multi-Armed Bandit (RMAB) problem \cite{guha2010approximation,xu2026smabsr}, where each arm corresponds to one persistent bottleneck flow with state $\Delta_m^e(t)$. At each micro-slot, the orchestrator selects $a_m(t)\in\{a_\varnothing\}\cup\mathcal{A}_m^+(t)$ for each pre-collapse bottleneck, while a bottleneck at $\Delta_{\mathrm{crit}}^e$ admits no active rescue and renews according to Eq.~(\ref{eq:collapse_renewal}).

Let $K$ denote the maximum number of new rescue attempts admitted per micro-slot. The global admission constraint is
\begin{equation} \label{eq:hard_constraint}
    \sum_{m \in \mathcal{M}}
    \mathbb{I}\left\{
    a_m(t)\neq a_\varnothing
    \right\}
    \leq K,
    \quad \forall t .
\end{equation}
Fidelity qualification is enforced through $\mathcal{A}_m^+(t)$, geometry and atmospheric conditions through $p_a(m,t)$, physical collapse through Eq.~(\ref{eq:collapse_renewal}), and local hardware occupancy by the asynchronous executor. Hence, Eq.~(\ref{eq:hard_constraint}) is the sole global coupling constraint.

Starlight seeks a non-anticipatory policy $\pi\in\Pi$ based on the current state $\mathbf{S}(t)$ and successful-completion probabilities $\{p_a(m,t)\}$. Rather than directly optimizing goodput or satisfaction within the reduced RMAB, we minimize long-term weighted EAoS as a tractable surrogate for maintaining physically viable rescue states under scarce capacity. With priority weight $\Omega_m$, the problem is
\begin{equation} \label{eq:objective}
    \mathcal{P}_1:
    \quad
    \min_{\pi \in \Pi}
    \limsup_{T \to \infty}
    \frac{1}{T}
    \mathbb{E}^{\pi}
    \left[
    \sum_{t=1}^{T}
    \sum_{m \in \mathcal{M}}
    \Omega_m \Delta_m^e(t)
    \right]
\end{equation}
\begin{equation*}
    \text{s.t.}
    \quad
    \text{Eq.~(\ref{eq:hard_constraint})}.
\end{equation*}
Here, $\mathbb{E}^{\pi}[\cdot]$ is taken over the regenerative dynamics in Eqs.~(\ref{eq:transition_dynamics}) and~(\ref{eq:collapse_renewal}); fidelity-qualified goodput, satisfaction, and instance collapse ratio are evaluated separately as system-level outcomes.

\section{Whittle-Driven Rescue Orchestration}
\label{sec:whittle}
\subsection{Lagrangian Relaxation and Decoupling}

We relax the instantaneous admission constraint in Eq.~(\ref{eq:hard_constraint}) to its time-average counterpart and charge a nonnegative intervention tax $W\geq0$ for each active rescue. The relaxed per-slot cost of bottleneck $m$ is
\begin{equation} \label{eq:instantaneous_cost}
c_m(\Delta_m^e,a)
\triangleq
\Omega_m\Delta_m^e
+
W\mathbb{I}\{a\neq a_\varnothing\}.
\end{equation}

The resulting relaxed problem is
\begin{equation} \label{eq:relaxed_problem}
\min_{\pi\in\Pi}
\limsup_{T\to\infty}
\frac{1}{T}
\mathbb{E}^{\pi}
\left[
\sum_{t=1}^{T}
\sum_{m\in\mathcal{M}}
c_m\bigl(\Delta_m^e(t),a_m(t)\bigr)
\right]
-WK.
\end{equation}

Absorbing the shared admission coupling into $W$ decomposes Eq.~(\ref{eq:relaxed_problem}) into $|\mathcal{M}|$ independent single-bottleneck subproblems, while the state-dependent fidelity feasibility $\mathcal{A}_m^+(t)$ remains local. For bottleneck $m$, the relaxed objective is
\begin{equation} \label{eq:single_link_objective}
\min_{\pi_m}
\limsup_{T\to\infty}
\frac{1}{T}
\mathbb{E}^{\pi_m}
\left[
\sum_{t=1}^{T}
c_m\bigl(\Delta_m^e(t),a_m(t)\bigr)
\right].
\end{equation}

\subsection{Single-Bottleneck Bellman Characterization}

We consider one bottleneck and omit subscript $m$. Under a fixed physical profile, the drift $\Gamma$, action parameters $\{p_a,\psi_a\}$, and renewal distribution $\mathcal{D}$ are stationary and exogenous to the current EAoS state. Let $\theta$ and $V(\Delta^e)$ denote the optimal average cost per micro-slot and differential value function, respectively, with $\Delta^{e+}\triangleq\phi(\Delta^e)$ denoting the one-step passive successor state.

For a pre-collapse state $\Delta^e<\Delta_{\mathrm{crit}}^e$, the admissible action set is $\mathcal{A}(\Delta^e)\triangleq\{a_\varnothing\}\cup\mathcal{A}^+(\Delta^e)$. The corresponding average-cost Bellman equation \cite{puterman2014markov} is
\begin{equation} \label{eq:bellman}
V(\Delta^e)+\theta
=
\min_{a\in\mathcal{A}(\Delta^e)}
\left\{
c(\Delta^e,a)
+
\mathbb{E}\!\left[
V(\Delta^{e'})
\mid
\Delta^e,a
\right]
\right\}.
\end{equation}

The collapse state is terminal rather than a decision state: reaching $\Delta_{\mathrm{crit}}^e$ terminates the current instance, incurs its terminal state cost, and renews the bottleneck to $\Delta_0^e\sim\mathcal{D}$ according to Eq.~(\ref{eq:collapse_renewal}). Hence,
\begin{equation} \label{eq:collapse_bellman}
V(\Delta_{\mathrm{crit}}^e)+\theta
=
\Omega\Delta_{\mathrm{crit}}^e
+
\mathbb{E}_{\mathcal{D}}
\!\left[
V(\Delta_0^e)
\right].
\end{equation}

\textbf{Passive Idling ($a_\varnothing$):} For a pre-collapse state, passive idling incurs no intervention tax and allows the current instance to follow its deterministic EAoS drift:
\begin{equation} \label{eq:q_passive}
Q(\Delta^e,a_\varnothing)
=
\Omega\Delta^e
+
V(\Delta^{e+}).
\end{equation}

\textbf{Active Intervention ($a\in\mathcal{A}^+(\Delta^e)$):} An active rescue incurs the intervention tax $W$. With successful-completion probability $p_a$, the current instance terminates and renews to $\Delta_0^e\sim\mathcal{D}$; otherwise, the same unresolved instance continues to $\Delta^{e+}$. Thus,
\begin{equation} \label{eq:q_active}
Q(\Delta^e,a)
=
\Omega\Delta^e
+
W
+
p_a
\mathbb{E}_{\mathcal{D}}
\!\left[
V(\Delta_0^e)
\right]
+
(1-p_a)V(\Delta^{e+}).
\end{equation}

For a fixed fidelity-feasible mode $a$, let $W^*(\Delta^e,a)$ denote an intervention tax for which active rescue and passive idling are equally preferable:
\begin{equation} \label{eq:indifference}
Q(\Delta^e,a)
=
Q(\Delta^e,a_\varnothing).
\end{equation}
Substituting Eqs.~(\ref{eq:q_passive}) and~(\ref{eq:q_active}) yields
\begin{equation} \label{eq:w_isolated}
W^*(\Delta^e,a)
=
p_a
\left[
V_{W^*}(\Delta^{e+})
-
\mathbb{E}_{\mathcal{D}}
\!\left[
V_{W^*}(\Delta_0^e)
\right]
\right].
\end{equation}
Eq.~(\ref{eq:w_isolated}) is an exact Bellman indifference relation for the fixed-mode, completion-embedded single-bottleneck subproblem. Evaluating it directly, however, requires the differential value function and is therefore unsuitable for micro-slot orchestration.

\subsection{Closed-Form Whittle-Style Rescue Score}

To obtain a lightweight approximation, we consider the fidelity-feasible state region of a fixed active mode $a$,
\begin{equation} \label{eq:feasible_state_region}
\mathcal{X}_a^+
\triangleq
\left\{
\Delta^e\in[0,\Delta_{\mathrm{crit}}^e):
\Delta^e+\psi_a\leq\Delta_{\mathrm{req}}^e
\right\}.
\end{equation}
For the fluid approximation below, we impose the one-step non-crossing condition
\begin{equation} \label{eq:non_crossing}
\phi(\Delta^e)
<
\Delta_{\mathrm{crit}}^e,
\qquad
\forall\,\Delta^e\in\mathcal{X}_a^+ .
\end{equation}
A sufficient condition is $\Delta_{\mathrm{req}}^e-\psi_a+\Gamma\delta t<\Delta_{\mathrm{crit}}^e$, which keeps all score-derivation successors within the pre-collapse region and separates the fluid approximation from the terminal-renewal boundary in Eq.~(\ref{eq:collapse_bellman}).

Under Eq.~(\ref{eq:non_crossing}), passive EAoS evolves locally as the linear drift $\Delta^{e+}=\Delta^e+\Gamma\delta t$. Motivated by the area-under-age construction commonly used in age-based scheduling \cite{weber1991addendum,hsu2018age,yates2021age}, we approximate the differential value accumulated under this drift by the quadratic fluid surrogate
\begin{equation} \label{eq:fluid_value}
V(x)
\approx
\frac{\Omega}{2\Gamma\delta t}x^2.
\end{equation}
Substituting Eq.~(\ref{eq:fluid_value}) into the Bellman indifference relation in Eq.~(\ref{eq:w_isolated}) gives the action-specific Whittle-style rescue score
\begin{equation} \label{eq:closed_form_action}
\widehat{W}(\Delta^e,a)
=
\frac{p_a\Omega}{2\Gamma\delta t}
\left[
(\Delta^{e+})^2
-
\mathbb{E}_{\mathcal{D}}
\!\left[
(\Delta_0^e)^2
\right]
\right].
\end{equation}

\begin{proposition}[Monotonicity of the closed-form rescue score]
\label{prop:score_monotonicity}
Consider a fixed active rescue mode $a$ under a stationary physical profile with $\Gamma>0$, $\Omega>0$, $p_a\geq0$, and state-independent renewal distribution $\mathcal{D}$. Under the one-step non-crossing condition in Eq.~(\ref{eq:non_crossing}), $\widehat{W}(\Delta^e,a)$ is nondecreasing in $\Delta^e$ over $\mathcal{X}_a^+$, and is strictly increasing whenever $p_a>0$.
\end{proposition}

\emph{Proof.} Over $\mathcal{X}_a^+$, Eq.~(\ref{eq:non_crossing}) gives $\Delta^{e+}=\Delta^e+\Gamma\delta t$. All remaining terms in Eq.~(\ref{eq:closed_form_action}) are constant with respect to $\Delta^e$, and
\begin{equation}
\frac{\partial\widehat{W}(\Delta^e,a)}
{\partial\Delta^e}
=
\frac{p_a\Omega}{\Gamma\delta t}
\left(
\Delta^e+\Gamma\delta t
\right)
\geq0.
\end{equation}
The inequality is strict for $p_a>0$, completing the proof.

Restoring the bottleneck index and the current physical profile, Starlight evaluates Eq.~(\ref{eq:closed_form_action}) over the current fidelity-feasible action set $\mathcal{A}_m^+(t)$. The resulting bottleneck-level rescue index is
\begin{equation} \label{eq:final_index}
I_m(t)
=
\max_{a\in\mathcal{A}_m^+(t)}
\widehat{W}_m
\!\left(
\Delta_m^e(t),a
\right),
\end{equation}
with $I_m(t)=0$ when $\mathcal{A}_m^+(t)=\emptyset$. The corresponding rescue mode is
\begin{equation} \label{eq:selected_action}
a_m^*(t)
=
\arg\max_{a\in\mathcal{A}_m^+(t)}
\widehat{W}_m
\!\left(
\Delta_m^e(t),a
\right).
\end{equation}

Because feasible modes of a bottleneck share the same state-dependent factor in Eq.~(\ref{eq:closed_form_action}), a positive-index bottleneck selects the fidelity-feasible action with the largest current successful-completion probability $p_a(m,t)$; if this common factor is nonpositive, no active mode is admitted. Thus, fidelity determines action admissibility, the current physical profile selects among feasible modes, and the Whittle-style score determines inter-bottleneck admission priority.

The score in Eq.~(\ref{eq:closed_form_action}) is a Bellman-derived fluid approximation rather than the exact Whittle index of the full state-dependent multi-action arm. Proposition~\ref{prop:score_monotonicity} establishes its monotonic urgency ordering under the stated fixed-profile conditions, while Section~\ref{subsec:triage_efficiency} evaluates its decision quality against exact value iteration.

\section{Starlight Protocol Implementation}
\label{sec:implementation}
\subsection{Decoupled Two-Timescale Control Plane}
\label{subsec:two_timescale}

\begin{figure}[t]
    \centering
    \includegraphics[width=0.95\columnwidth]{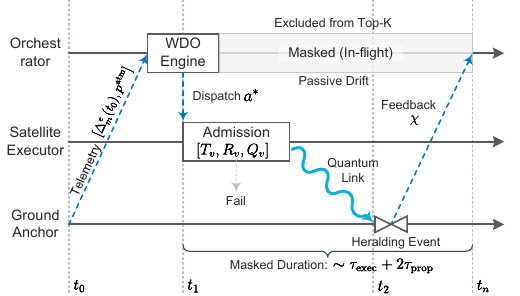}
    \caption{\textbf{Asynchronous execution pipeline.} The \textit{Masked} state spans action-specific physical execution and the round-trip heralding blind spot ($2\tau_{\mathrm{prop}}$, where $\tau_{\mathrm{prop}}$ is the one-way classical propagation delay), preventing repeated admission of the same in-flight instance until successful completion, failure/timeout, or collapse.}
    \label{fig:protocol_pipeline}
\end{figure}

Starlight employs a logically centralized control plane in which terrestrial Anchors report EAoS states and atmospheric telemetry to an orchestrator that runs WDO and enforces the global admission limit $K$. To amortize geometry and physical-realization planning while retaining fast adaptation, control is decoupled into macro-epochs ($\Delta T$) and micro-slots ($\delta t$).

At each macro-epoch, the orchestrator evaluates geometry-dependent feasibility and retains a preferred single-satellite realization for each feasible bottleneck--mode pair. The realization must provide simultaneous dual-Anchor coverage for $a_{\parallel}$ or feasible asynchronous contacts for $a_{\bowtie}$, and determines the slowly varying component $p_a^{\mathrm{geom}}(m,t)$ and, for $a_{\bowtie}$, the synchronization delay $\bar{\tau}_m$. The successful-completion probability is factorized as
\begin{equation} \label{eq:prob_decoupling}
    p_a(m,t)
    =
    p_a^{\mathrm{geom}}(m,t)
    \cdot
    p_a^{\mathrm{atm}}(m,t).
\end{equation}

Within each macro-epoch, $p_a^{\mathrm{geom}}$ remains fixed and equals zero when no feasible realization exists, whereas $p_a^{\mathrm{atm}}$ is refreshed every micro-slot from Anchor atmospheric telemetry. Local hardware resources are checked only after dispatch by the selected executor.

\subsection{Whittle-Driven Micro-Slot Orchestration}
\label{subsec:wdo_orchestration}

At each micro-slot, the orchestrator updates $p_a(m,t)$ using the latest atmospheric telemetry and the current macro-epoch geometry profile. For each unmasked bottleneck, WDO constructs $\mathcal{A}_m^+(t)$, selects the feasible mode maximizing Eq.~(\ref{eq:closed_form_action}), and ranks the bottleneck by $I_m(t)$. The top-$K$ positive-index bottlenecks are dispatched to their retained satellite realizations, as summarized in Algorithm~\ref{alg:wdo}.

Because rescue execution may span multiple micro-slots, WDO masks an instance once its dispatch is accepted. The masked instance continues passive EAoS evolution but is excluded from further arbitration until completion, failure/timeout, or collapse, as illustrated in Fig.~\ref{fig:protocol_pipeline}.

\renewcommand{\algorithmicrequire}{\textbf{Input:}}
\renewcommand{\algorithmicensure}{\textbf{Output:}}
\begin{algorithm}[t]
\caption{Whittle-Driven Orchestration (WDO)}
\label{alg:wdo}
\begin{algorithmic}[1]
\Require States $\{\Delta_m^e(t)\}$, budget $K$, probabilities $\{p_a(m,t)\}$
\Ensure Dispatch set $\mathcal{R}(t)=\{(m,a)\}$ (up to $K$)

\State $\mathcal{L}\gets\emptyset$
\For{each bottleneck $m\in\mathcal{M}$}
    \If{\textsc{Masked}$(m)$}
        \State \textbf{continue}
    \EndIf
    \State Construct $\mathcal{A}_m^+(t)$ using Eq.~(\ref{eq:action_feasibility})
    \If{$\mathcal{A}_m^+(t)=\emptyset$}
        \State \textbf{continue}
    \EndIf
    \State $a_m^*(t)\gets\arg\max_{a\in\mathcal{A}_m^+(t)}
    \widehat{W}_m(\Delta_m^e(t),a)$
    \State $I_m(t)\gets\widehat{W}_m(\Delta_m^e(t),a_m^*(t))$
    \If{$I_m(t)>0$}
        \State Append $(m,a_m^*(t),I_m(t))$ to $\mathcal{L}$
    \EndIf
\EndFor
\State $\mathcal{R}(t)\gets$ top-$K$ entries of $\mathcal{L}$ ranked by $I_m(t)$
\State Dispatch $\mathcal{R}(t)$ to the selected satellite realizations
\end{algorithmic}
\end{algorithm}
\subsection{Physical Execution of Rescue Actions}
\label{subsec:physical_execution}

While WDO determines \emph{what} to admit, the selected satellite executor determines \emph{how} to realize the dispatched mode under instantaneous hardware constraints. Upon receiving $(m,a_m^*(t))$, it checks the required satellite resources and Anchor receivers; for $a_{\bowtie}$, the first satellite--Anchor entanglement is buffered until the second becomes available for the in-orbit BSM.

If hardware admission fails, the attempt is suppressed and the instance follows the passive transition in Eq.~(\ref{eq:transition_dynamics}). Otherwise, the executor reserves the required resources, starts the rescue, and associates delayed feedback with the current renewal generation. Successful completion ($\chi=1$) releases resources and renews the instance to $\Delta_{m,0}^e\sim\mathcal{D}_m$, whereas failure or timeout ($\chi=0$) releases resources and leaves the instance on its passive trajectory.

If collapse occurs while a rescue remains pending, the current instance renews according to Eq.~(\ref{eq:collapse_renewal}), the pending rescue is invalidated, and any subsequent feedback is discarded as stale.

\subsection{Computational Complexity}
\label{subsec:complexity_overhead}

Given the macro-epoch geometry profile, updating constant-size action profiles for all bottlenecks requires $\mathcal{O}(|\mathcal{M}|)$ work per macro-epoch. At each micro-slot, WDO evaluates a constant number of closed-form scores per bottleneck and extracts the top-$K$ positive indices with a size-$K$ min-heap, yielding $\mathcal{O}(|\mathcal{M}|\log K)$ orchestration time. Executor-side admission, masking, validation, and resource bookkeeping incur $\mathcal{O}(1)$ overhead per dispatched rescue.

\section{Performance Evaluation}
\label{sec:evaluation}
\subsection{Methodology}
\label{subsec:methodology}
\noindent\textbf{Simulation setup.} We implement Starlight in QUASAR \cite{shi2026quasar}, our lightweight event-driven simulator for satellite quantum networks built on the SimQN framework \cite{chen2023simqn}. The simulator co-simulates Walker-delta orbital dynamics, visibility-gated space-to-ground links, and micro-slot EAoS evolution using the default parameters in Table~\ref{tab:params}. We sample 128 inter-domain bottlenecks from distinct cross-domain Anchor pairs and keep them fixed across all policies. Geometry-aware bindings refresh every macro-epoch, while WDO operates per micro-slot. Successful-completion probabilities $p_a(m,t)$ combine geometric feasibility with atmospheric variation modeled by Gilbert--Elliott cloud outages and temporally correlated log-normal scintillation \cite{yehia2024connecting,khatri2021spooky}. Space-to-ground transmissivity is abstracted by a 1~kHz clear-sky generation rate per terminal, modulated by geometry and atmospheric conditions. Satellites provide $T_v\in[6,10]$ transmitters and $Q_v\in[16,64]$ quantum memories, while Anchors provide $R_u\in[2,6]$ receivers. Results are averaged over 20 independent seeds.

\noindent\textbf{Workload and state evolution.} Each bottleneck $m\in\mathcal{M}$ carries persistent backlogged entanglement demand with target fidelity-qualified service rate $d=4$\,pairs/s, where $d$ defines the satisfaction target rather than a goodput cap. The satellite-rescue abstraction absorbs the latency of preparing subsequent intra-domain subpaths into the instance-exposure process. Each newly exposed ready-but-unbridged instance starts with source fidelity $F_0=0.99$, corresponding to $\lambda_0=(4F_0-1)/3$, and renewal state $\Delta_{m,0}^e\sim\mathrm{Unif}[0.1,0.3]$.

Buffered states decay at terrestrial ($\gamma_{\mathrm{G}}=1.0\,\mathrm{s}^{-1}$) or space-borne ($\gamma_{\mathrm{S}}=0.1\,\mathrm{s}^{-1}$) rates \cite{huang2024vacuum}. Following standard swapping-penalty abstractions \cite{popa2024skyswapping}, we set $P_{\mathrm{G}}=P_{\mathrm{S}}=0.05$ for ground and in-orbit BSMs. We set $\Delta_{\mathrm{req}}^e=1.0$, $\Delta_{\mathrm{crit}}^e=1.5$, and $\Omega_m=1$ unless otherwise specified.

For the memory-heterogeneity experiment in Fig.~\ref{fig:memory_heterogeneity}, bottlenecks are evenly divided into slow- and fast-decay groups independently of channel quality and action feasibility. We vary $r_{\gamma}=\gamma_{\mathrm{fast}}/\gamma_{\mathrm{slow}}$ from 1 to 5 while fixing the mean terrestrial decay rate at $\gamma_{\mathrm{G}}=1.0\,\mathrm{s}^{-1}$, giving $\gamma_{\mathrm{slow}}=2\gamma_{\mathrm{G}}/(1+r_{\gamma})$ and $\gamma_{\mathrm{fast}}=2r_{\gamma}\gamma_{\mathrm{G}}/(1+r_{\gamma})$.
\begin{table}[t]
\centering
\caption{Default simulation parameters.}
\label{tab:params}
\footnotesize
\setlength{\tabcolsep}{2.7pt}
\renewcommand{\arraystretch}{1.02}
\begin{tabular}{@{}lr@{\hspace{8pt}}lr@{}}
\toprule
\textbf{Parameter} & \textbf{Value} & \textbf{Parameter} & \textbf{Value} \\
\midrule
Domains $D$ & 12 &
Satellites $N$ & 48 \\

Anchors & 24 &
Planes $\times$ Sats/plane & $6\times8$ \\

Bottlenecks $|\mathcal{M}|$ & 128 &
Walker phasing $F$ & 1 \\

Orbital altitude $h$ & 300 km &
Inclination $i$ & $53^\circ$ \\

Orbital period $T_{\mathrm{orb}}$ & 90 min &
Elevation mask & $20^\circ$ \\

Simulation horizon $T_{\mathrm{sim}}$ & 5400 s &
Warm-up $T_{\mathrm{warm}}$ & 60 s \\

Macro-epoch $\Delta T$ & 30 s &
Micro-slot $\delta t$ & 10 ms \\

Random seeds & 20 &
Cloud outage $p_{\mathrm{cloud}}$ & 0.3 \\

Clear duration $\bar{T}_{\mathrm{clear}}$ & 70 s &
Outage duration $\bar{T}_{\mathrm{out}}$ & 30 s \\

Scint. index $\sigma_I^2$ & 0.10 &
Scint. corr. time $\tau_{\mathrm{scin}}$ & 100 ms \\

Target rate $d$ & 4 pairs/s &
Admission budget $K$ & 6 \\
\bottomrule
\end{tabular}
\end{table}

\noindent\textbf{Baselines.} We compare Starlight with representative planning-based, age-driven, state-reactive, and channel-greedy policies.
\begin{itemize}
    \item \textbf{TE-MILP:} A periodically replanned time-expanded MILP adapted from prior entanglement scheduling formulations~\cite{panigrahy2022optimal,chang2024entanglement,wei2024optimal}. It maximizes expected fidelity-qualified E2E pair delivery under the same system constraints, with the schedule fixed within each macro-epoch.
    \item \textbf{AoE-Greedy:} An age-driven heuristic inspired by AoE and Fidelity-Age scheduling~\cite{ceran2026age,ercetin2026fidelity}, ranking bottlenecks by the elapsed time since instance exposure.
    \item \textbf{EAoS-Myopic:} A micro-slot reactive policy ranking bottlenecks by instantaneous $\Delta_m^e(t)$, isolating the forward-looking benefit of the Whittle-style score.
    \item \textbf{Max-Feasibility:} A channel-greedy heuristic ranking bottlenecks by $\max_{a\in\mathcal{A}_m^+(t)}p_a(m,t)$, without accounting for rescue urgency.
\end{itemize}

All policies use the same target rate $d$, admission budget $K$, hardware limits, and stochastic environmental traces. TE-MILP uses a logical time-expanded graph to encode action feasibility and waiting-induced storage loss. For the online heuristics, each policy retains its native ranking metric, while the selected bottleneck uses the fidelity-feasible mode with the largest current $p_a(m,t)$, without oracle information.

\noindent\textbf{Evaluation Metrics.} We evaluate three performance dimensions:
\begin{itemize}
    \item \textbf{E2E Goodput and Average Satisfaction Ratio:} Let $g_m$ denote the fidelity-qualified E2E delivery rate of bottleneck $m$. Network goodput is $G=\sum_{m\in\mathcal{M}} g_m$ pairs/s. Its satisfaction ratio is $s_m=\min\{g_m/d,1\}$, and the network average is $S=|\mathcal{M}|^{-1}\sum_{m\in\mathcal{M}}s_m$.
    \item \textbf{Average EAoS and Collapse Ratio:} Network EAoS is $(T|\mathcal{M}|)^{-1}\sum_{t,m}\Omega_m\Delta_m^e(t)$; the instance collapse ratio is $C_{\mathrm{col}}=N_{\mathrm{col}}/(N_{\mathrm{succ}}+N_{\mathrm{col}})$, where $N_{\mathrm{col}}$ and $N_{\mathrm{succ}}$ count collapsed and successfully rescued instances.
    \item \textbf{Orchestration Time:} Wall-clock time required to produce one scheduling decision; for Starlight, this includes score evaluation and top-$K$ selection.
\end{itemize}
\subsection{Resource Efficiency}
\label{subsec:triage_efficiency}
\begin{figure*}[!t]
\centering

\begin{minipage}[t]{0.485\textwidth}
\centering
\includegraphics[width=\linewidth]{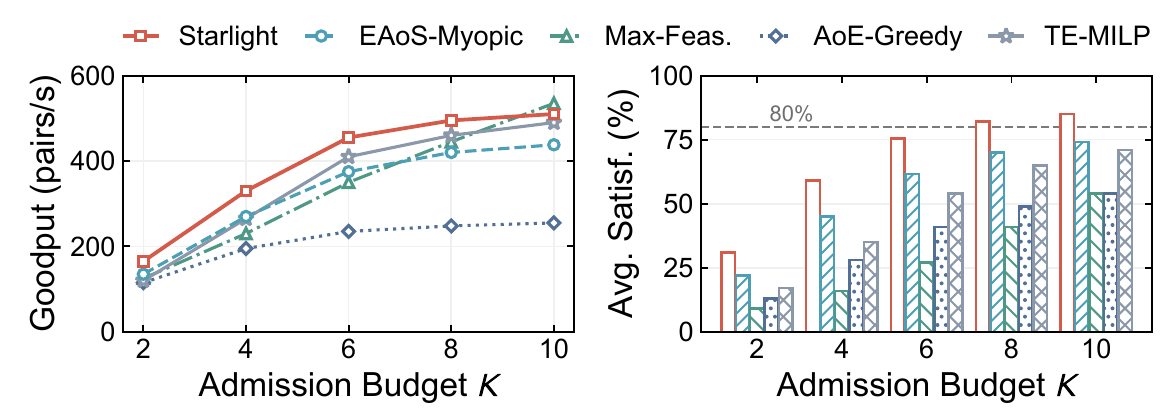}

\begin{minipage}{0.49\linewidth}
\centering
{\footnotesize (a) E2E Goodput}
\end{minipage}\hfill
\begin{minipage}{0.49\linewidth}
\centering
{\footnotesize (b) Avg. Satisfaction}
\end{minipage}

\vspace{-0.25cm}
\caption{Experimental results under resource starvation.}
\label{fig:capacity}
\end{minipage}
\hfill
\begin{minipage}[t]{0.485\textwidth}
\centering
\includegraphics[width=\linewidth]{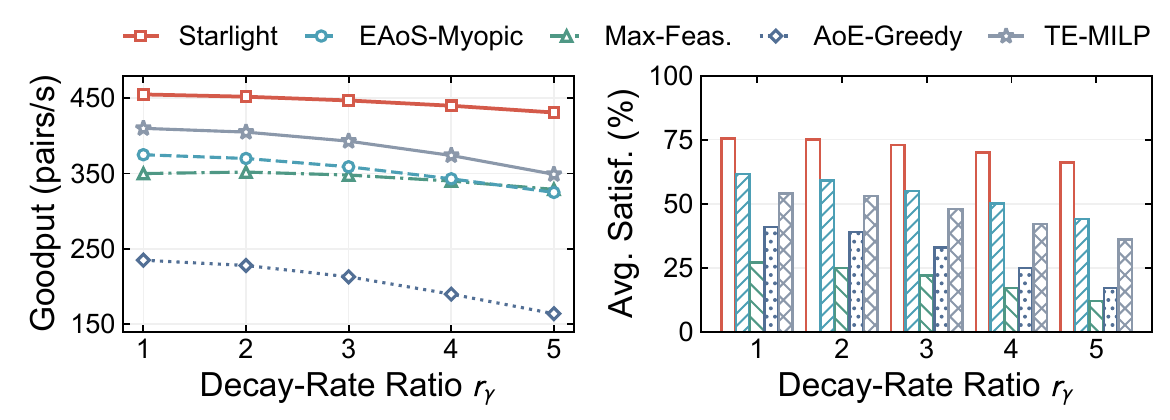}

\begin{minipage}{0.49\linewidth}
\centering
{\footnotesize (a) E2E Goodput}
\end{minipage}\hfill
\begin{minipage}{0.49\linewidth}
\centering
{\footnotesize (b) Avg. Satisfaction}
\end{minipage}

\vspace{-0.25cm}
\caption{Experimental results under memory heterogeneity.}
\label{fig:memory_heterogeneity}
\end{minipage}

\vspace{-0.15cm}
\end{figure*}
\begin{figure*}[!t]
\centering

\begin{minipage}[t]{0.485\textwidth}
\centering
\includegraphics[width=\linewidth]{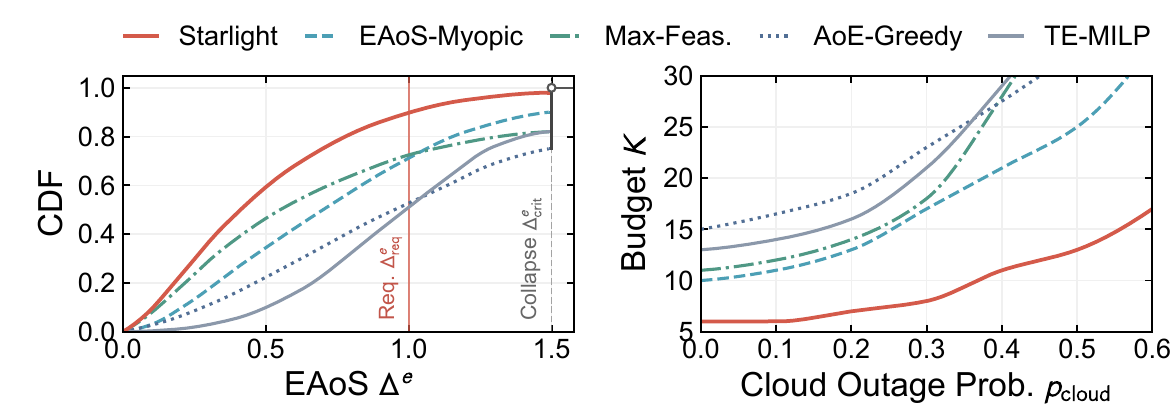}

\begin{minipage}{0.49\linewidth}
\centering
{\footnotesize (a) EAoS CDF}
\end{minipage}\hfill
\begin{minipage}{0.49\linewidth}
\centering
{\footnotesize (b) Survival Frontier}
\end{minipage}

\vspace{-0.25cm}
\caption{Atmospheric resilience and survival frontier.}
\label{fig:atmospheric_resilience}
\end{minipage}
\hfill
\begin{minipage}[t]{0.485\textwidth}
\centering
\includegraphics[width=\linewidth]{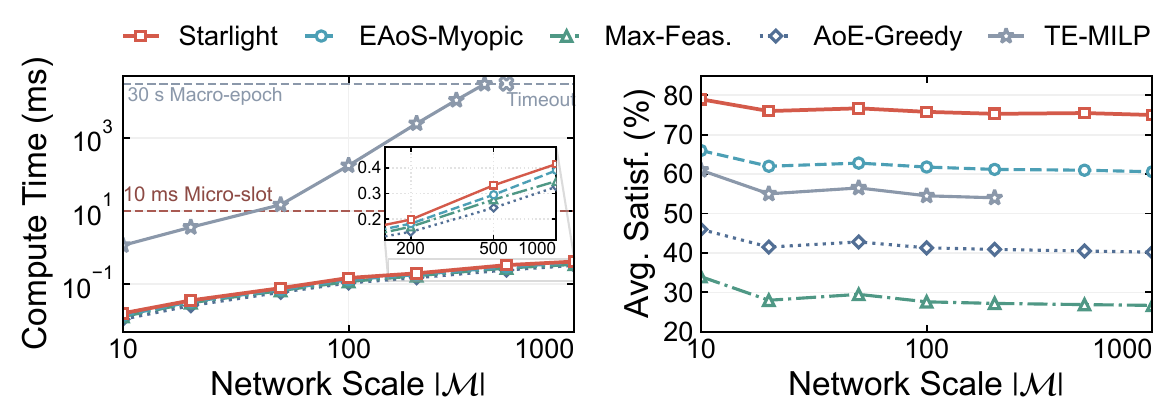}

\begin{minipage}{0.49\linewidth}
\centering
{\footnotesize (a) Compute Time}
\end{minipage}\hfill
\begin{minipage}{0.49\linewidth}
\centering
{\footnotesize (b) Avg. Satisfaction}
\end{minipage}

\vspace{-0.25cm}
\caption{Scalability under increasing network size.}
\label{fig:scalability}
\end{minipage}

\vspace{-0.15cm}
\end{figure*}

\noindent\textbf{Index Approximation Validation.} Because exact Value Iteration (VI) scales exponentially with the number of bottlenecks, we use a tractable micro-topology with $|\mathcal{M}|=5$ and $K=2$. Each bottleneck EAoS is discretized over $[0,\Delta_{\mathrm{crit}}^e]$ with interval $\delta_{\mathrm{VI}}=0.15$, yielding 11 states, with linear interpolation for off-grid successors. Starlight and VI use the same state grid, transition model, successful-completion probabilities, renewal distribution, and admission budget. Normalizing Eq.~(\ref{eq:objective}) by $|\mathcal{M}|$, exact VI achieves a per-bottleneck average cost of $0.317$, versus $0.332$ for Starlight, corresponding to a $4.73\%$ gap. This result indicates that the fluid approximation closely tracks the exact decision quality without online Bellman-equation computation.

\noindent\textbf{Resource Starvation.} We evaluate the default topology with $|\mathcal{M}|=128$ while varying the admission budget as $K\in\{2,4,6,8,10\}$. As shown in Fig.~\ref{fig:capacity}(a), Starlight's E2E goodput increases from $165$ pairs/s at $K=2$ to $455$ pairs/s at the default $K=6$, and gradually saturates at $510$ pairs/s for $K=10$. EAoS-Myopic follows a similar but lower trajectory, while AoE-Greedy saturates earlier as chronological age becomes less informative of physical degradation. TE-MILP remains competitive as capacity increases, but its macro-epoch-fixed schedule limits adaptation to rapidly changing states. Max-Feasibility instead reaches $535$ pairs/s at $K=10$, exceeding Starlight by preferentially exploiting highly feasible channel opportunities.

Fig.~\ref{fig:capacity}(b) shows that this aggregate-goodput advantage does not translate into balanced service. At $K=10$, Max-Feasibility achieves only $54.0\%$ average satisfaction, compared with $84.5\%$ for Starlight, because channel-greedy allocation repeatedly favors bottlenecks with favorable instantaneous feasibility. EAoS-Myopic and TE-MILP reach $74.0\%$ and $71.0\%$, respectively. Starlight maintains the highest satisfaction across the evaluated budgets, indicating that its advantage under resource scarcity lies in allocating rescue opportunities according to physical urgency rather than maximizing aggregate goodput alone.
\subsection{Physical Heterogeneity and Uncertainty}
\label{subsec:resilience}
\noindent\textbf{Memory Heterogeneity.} We next examine robustness to increasingly heterogeneous terrestrial-memory decay. As defined in Section~\ref{subsec:methodology}, Fig.~\ref{fig:memory_heterogeneity} varies $r_{\gamma}=\gamma_{\mathrm{fast}}/\gamma_{\mathrm{slow}}$ from 1 to 5 while keeping the mean terrestrial decay rate fixed. At $r_{\gamma}=1$, the results recover the homogeneous-memory operating point at the default $K=6$. As heterogeneity increases, Starlight degrades moderately, with goodput decreasing from $455$ to $431$ pairs/s and average satisfaction from $75.5\%$ to $66.0\%$. EAoS-Myopic is more sensitive, falling from $375$ to $325$ pairs/s and from $61.5\%$ to $44.0\%$ satisfaction, while AoE-Greedy drops from $235$ to $164$ pairs/s and from $41.0\%$ to $17.0\%$. The stronger degradation of AoE-Greedy reflects the growing mismatch between chronological age and physical degradation as memory decay becomes more heterogeneous.

Max-Feasibility exhibits a different trade-off: its goodput decreases only from $350$ to $329$ pairs/s, while satisfaction falls from $27.0\%$ to $12.0\%$, indicating that channel-greedy scheduling can preserve aggregate delivery by repeatedly favoring highly feasible links while neglecting faster-decaying bottlenecks. TE-MILP decreases from $410$ to $349$ pairs/s and from $54.0\%$ to $36.0\%$ satisfaction, consistent with the limited responsiveness of its macro-epoch-fixed schedule to heterogeneous EAoS evolution. Overall, Starlight's growing advantage under larger $r_{\gamma}$ highlights the importance of decay-aware state representation, while its separation from EAoS-Myopic further reflects the benefit of forward-looking prioritization.

\noindent\textbf{Atmospheric Robustness.} Fig.~\ref{fig:atmospheric_resilience}(a) reports the bottleneck-EAoS distribution under the default $p_{\mathrm{cloud}}=0.3$ and $K=6$. Starlight shifts the distribution toward the low-EAoS region, with approximately $90\%$ of bottleneck states below $\Delta_{\mathrm{req}}^e=1.0$ and only a small tail approaching $\Delta_{\mathrm{crit}}^e=1.5$. The competing policies exhibit heavier tails, indicating less effective prioritization under stochastic atmospheric degradation.

Fig.~\ref{fig:atmospheric_resilience}(b) reports the minimum admission budget required to sustain $80\%$ average satisfaction as $p_{\mathrm{cloud}}$ increases. Starlight requires $K=6$ under clear conditions, $K=8$ at the default $p_{\mathrm{cloud}}=0.3$, and $K=17$ at $p_{\mathrm{cloud}}=0.6$. At $p_{\mathrm{cloud}}=0.3$, EAoS-Myopic, Max-Feasibility, AoE-Greedy, and TE-MILP require $K=17$, $18$, $23$, and $21$, respectively. Their frontiers rise more rapidly as outages intensify and eventually exceed the displayed budget range, whereas Starlight remains within it. Thus, Starlight does not eliminate atmospheric loss, but uses available rescue capacity more effectively through EAoS-aware and feasibility-sensitive prioritization.

\subsection{Scalability and Ablation}
\label{subsec:scalability_ablation}
\noindent\textbf{Scalability.} Fig.~\ref{fig:scalability}(a) reports wall-clock orchestration time as $|\mathcal{M}|$ increases from $10$ to $1{,}000$, with the dashed 10\,ms line marking the micro-slot timescale. Starlight remains well below this limit, increasing from $0.016$\,ms at $|\mathcal{M}|=10$ to $0.415$\,ms at $|\mathcal{M}|=1{,}000$, while the other online heuristics remain similarly sub-millisecond. TE-MILP requires $14.8$\,ms at $50$ bottlenecks and $2.45$\,s at $200$, and reaches the 30-s macro-epoch limit at $|\mathcal{M}|=500$. These results are consistent with Starlight's closed-form score evaluation and $\mathcal{O}(|\mathcal{M}|\log K)$ top-$K$ selection.

Fig.~\ref{fig:scalability}(b) evaluates scheduling effectiveness under increasing scale while setting $K=\max\{1,\lfloor0.05|\mathcal{M}|+0.5\rfloor\}$ to separate scaling effects from resource starvation. Starlight remains around $75\%$ satisfaction in the large-scale regime, compared with approximately $61\%$, $41\%$, and $27\%$ for EAoS-Myopic, AoE-Greedy, and Max-Feasibility, respectively. TE-MILP stays in the mid-$50\%$ range over tractable scales but is omitted once replanning reaches the macro-epoch limit. Overall, Starlight preserves both low orchestration overhead and stable scheduling effectiveness as the network scales.
\begin{table*}[t]
\centering
\vspace{-0.1cm}
\caption{Ablation study of Starlight components.}
\label{tab:ablation}
\setlength{\tabcolsep}{10pt}
\renewcommand{\arraystretch}{1.15}
\small
\begin{tabular}{@{} l c c c c c @{}}
\toprule
\textbf{Variant}
& \makecell{\textbf{Goodput}\\[-0.2ex]\scriptsize(pairs/s)}
& \makecell{\textbf{Avg. Sat.}\\[-0.2ex]\scriptsize(\%)}
& \makecell{\textbf{Collapse Ratio}\\[-0.2ex]\scriptsize(\%)}
& \makecell{\textbf{Avg.}\\[-0.2ex]\textbf{EAoS}}
& \makecell{\textbf{Redund. Alloc.}\\[-0.2ex]\scriptsize(\%)} \\
\midrule

\textbf{Starlight}
& \textbf{455}
& \textbf{75.5}
& \textbf{2.0}
& \textbf{0.49}
& \textbf{0.0} \\

\textit{w/o} EAoS
& 328~\textcolor{gray}{\scriptsize($\downarrow$27.9\%)}
& 50.5~\textcolor{gray}{\scriptsize($\downarrow$33.1\%)}
& 15.5~\textcolor{gray}{\scriptsize($\uparrow$7.8$\times$)}
& 0.83~\textcolor{gray}{\scriptsize($\uparrow$69.4\%)}
& 0.0 \\

\textit{w/o} Whittle
& 375~\textcolor{gray}{\scriptsize($\downarrow$17.6\%)}
& 61.5~\textcolor{gray}{\scriptsize($\downarrow$18.5\%)}
& 9.2~\textcolor{gray}{\scriptsize($\uparrow$4.6$\times$)}
& 0.72~\textcolor{gray}{\scriptsize($\uparrow$46.9\%)}
& 0.0 \\

\textit{w/o} Masking
& 420~\textcolor{gray}{\scriptsize($\downarrow$7.7\%)}
& 68.8~\textcolor{gray}{\scriptsize($\downarrow$8.9\%)}
& 4.8~\textcolor{gray}{\scriptsize($\uparrow$2.4$\times$)}
& 0.56~\textcolor{gray}{\scriptsize($\uparrow$14.3\%)}
& 14.8~\textcolor{gray}{\scriptsize(+14.8 pp)} \\

\bottomrule
\end{tabular}
\vspace{-0.2cm}
\end{table*}

\noindent\textbf{Ablation Study.} Table~\ref{tab:ablation} isolates the contributions of EAoS, Whittle-driven prioritization, and asynchronous masking. The \textit{w/o} EAoS variant replaces EAoS with chronological instance age within the same Whittle-style framework, \textit{w/o} Whittle ranks feasible bottlenecks directly by current EAoS, and \textit{w/o} Masking keeps in-flight bottlenecks eligible for subsequent arbitration. Removing EAoS causes the largest degradation: goodput falls from $455$ to $328$ pairs/s and satisfaction from $75.5\%$ to $50.5\%$, while the collapse ratio rises from $2.0\%$ to $15.5\%$. Removing Whittle-driven prioritization yields $375$ pairs/s goodput and $61.5\%$ satisfaction, with a $9.2\%$ collapse ratio. These results distinguish the roles of the two components: EAoS provides degradation-aware state representation, while the Whittle-style score improves the allocation of scarce rescue opportunities according to future state consequences.

The \textit{w/o} Masking variant isolates the execution layer. We define the redundant-allocation ratio as the fraction of dispatches issued to bottlenecks with an unfinished rescue already in flight. Disabling masking raises this ratio from $0$ to $14.8\%$, while reducing goodput and satisfaction by $7.7\%$ and $8.9\%$, respectively. Its smaller changes in EAoS and collapse ratio relative to the other ablations indicate a different failure mechanism: repeated admission of unresolved in-flight instances wastes scarce rescue capacity. Thus, masking complements the scheduling logic by preventing redundant allocations during asynchronous execution.

\section{Conclusion and Future Work}
\label{sec:conclusion}

This paper presented Starlight, a demand-driven fidelity rescue framework for satellite--terrestrial quantum networks. Starlight introduces EAoS to map heterogeneous decoherence and swapping-induced fidelity penalties into a unified additive degradation state, and formulates multi-bottleneck rescue admission under shared satellite capacity as an RMAB. A closed-form Whittle-style index approximation enables $\mathcal{O}(|\mathcal{M}|\log K)$ top-$K$ selection, while the asynchronous dispatch-and-mask pipeline prevents redundant in-flight allocations.
Evaluations show that Starlight achieves the highest average satisfaction among the evaluated policies under constrained rescue capacity while retaining competitive goodput, reaching $84.5\%$ satisfaction at $K=10$. It remains robust to memory heterogeneity and stochastic atmospheric degradation, with orchestration time below $0.5$\,ms for 1,000 monitored bottlenecks. Ablation results further confirm the complementary roles of EAoS, Whittle-driven prioritization, and asynchronous masking. Future work will extend Starlight to multipartite GHZ distribution and integrate dynamic inter-satellite routing, atmospheric prediction, and on-board quantum error correction.


\begin{thebibliography}{99}

\bibitem{wehner2018quantum}
S.~Wehner, D.~Elkouss, and R.~Hanson, ``Quantum internet: A vision for the road ahead,'' \emph{Science}, vol. 362, no. 6412, p. eaam9288, 2018.

\bibitem{khalid2024integrated}
U.~Khalid, M.~S.~Ulum, M.~Z.~Win, and H.~Shin, ``Integrated satellite-ground variational quantum sensing networks,'' \emph{IEEE Communications Magazine}, vol.~62, no.~10, pp.~20--27, 2024.

\bibitem{yin2017satellite}
J.~Yin \emph{et~al.}, ``Satellite-based entanglement distribution over 1200 kilometers,'' \emph{Science}, vol. 356, no. 6343, pp. 1140--1144, 2017.

\bibitem{goswami2025satellites}
S.~Goswami \emph{et~al.}, ``Satellites promise global-scale quantum networks,'' \emph{Optica Quantum}, vol.~3, no.~6, pp.~590--605, 2025.

\bibitem{de2023satellite}
L.~de Forges de Parny \emph{et~al.}, ``Satellite-based quantum information networks: Use cases, architecture, and roadmap,'' \emph{Communications Physics}, vol.~6, no.~1, p.~12, 2023.

\bibitem{ren2017ground}
J.-G. Ren \emph{et~al.}, ``Ground-to-satellite quantum teleportation,'' \emph{Nature}, vol. 549, no. 7670, pp. 70--73, 2017.

\bibitem{liao2018satellite}
S.-K.~Liao \emph{et~al.}, ``Satellite-relayed intercontinental quantum network,'' \emph{Physical Review Letters}, vol.~120, no.~3, p.~030501, 2018.

\bibitem{jiang2025quantum}
W.~Jiang, Y.~Zhang, H.~Han, and J.~Mu, ``Quantum communication in self-organizing satellite networks: Challenges and opportunities,'' \emph{IEEE Communications Standards Magazine}, vol.~9, no.~3, pp.~15--25, 2025.

\bibitem{shi2026efficient}
Y.~Shi, Z.~Zhao, G.~Min, X.~Wang, and L.~Xu, ``Efficient entanglement routing for enhancing throughput in quantum networks,'' \emph{IEEE Internet Things J.}, early access, 2026, doi: 10.1109/JIOT.2026.3736579.

\bibitem{panigrahy2022optimal}
N.~K.~Panigrahy, P.~Dhara, D.~Towsley, S.~Guha, and L.~Tassiulas, ``Optimal entanglement distribution using satellite based quantum networks,'' in \emph{Proc. IEEE INFOCOM Workshops}, 2022, pp. 1--6.

\bibitem{wei2024optimal}
X.~Wei, J.~Liu, L.~Fan, Y.~Guo, Z.~Han, and Y.~Wang, ``Optimal entanglement distribution problem in satellite-based quantum networks,'' \emph{IEEE Network}, vol.~39, no.~1, pp.~97--103, 2025.

\bibitem{chang2024entanglement}
A.~Chang, Y.~Wan, G.~Xue, and A.~Sen, ``Entanglement distribution in satellite-based dynamic quantum networks,'' \emph{IEEE Network}, vol.~38, no.~1, pp. 79--86, 2024.

\bibitem{wei2024entanglement}
X.~Wei \emph{et~al.}, ``Entanglement from sky: Optimizing satellite-based entanglement distribution for quantum networks,'' \emph{IEEE/ACM Transactions on Networking}, vol.~32, no.~6, pp. 5295--5309, 2024.

\bibitem{ceran2026age}
E.~T. Ceran, ``Age of entanglement in satellite repeater chains with intermittent availability,'' in \emph{Proc. IEEE INFOCOM 2026--IEEE Conf. Comput. Commun.}, Tokyo, Japan, 2026, pp.~1--6.

\bibitem{ercetin2026fidelity}
O.~Ercetin and Z.~Gedik, ``Fidelity-age-aware scheduling in quantum repeater networks,'' in \emph{Proc. Int. Conf. Quantum Commun., Netw., Comput. (QCNC)}, Kobe, Japan, 2026, pp. 304--311.

\bibitem{gundougan2021proposal}
M.~G{\"u}ndo{\u{g}}an \emph{et~al.}, ``Proposal for space-borne quantum memories for global quantum networking,'' \emph{npj Quantum Information}, vol. 7, no. 1, p. 128, 2021.

\bibitem{panahi2025energy}
F.~H. Panahi, ``Energy-efficient decoherence-aware entanglement generation for quantum internet of things,'' \emph{IEEE Internet Things J.}, vol.~12, no.~20, pp.~42509--42516, 2025.

\bibitem{guha2010approximation}
S.~Guha, K.~Munagala, and P.~Shi, ``Approximation algorithms for restless bandit problems,'' \emph{Journal of the ACM}, vol.~58, no.~1, Art.~no.~3, Dec.~2010.

\bibitem{shi2020concurrent}
S.~Shi and C.~Qian, ``Concurrent entanglement routing for quantum networks: Model and designs,'' in \emph{Proc. ACM SIGCOMM}, 2020, pp.~62--75.

\bibitem{zhao2021redundant}
Y.~Zhao and C.~Qiao, ``Redundant entanglement provisioning and selection for throughput maximization in quantum networks,'' in \emph{Proc. IEEE INFOCOM}, 2021, pp.~1--10.

\bibitem{yang2023asynchronous}
L.~Yang, Y.~Zhao, L.~Huang, and C.~Qiao, ``Asynchronous entanglement provisioning and routing for distributed quantum computing,'' in \emph{Proc. IEEE INFOCOM}, 2023, pp.~1--10.

\bibitem{kozlowski2024quip}
W.~Kozlowski, F.~Kuipers, R.~Smets, and B.~Turkovic, ``QuIP: A P4 quantum internet protocol prototyping framework,'' \emph{IEEE Journal on Selected Areas in Communications}, vol.~42, no.~7, pp.~1936--1949, 2024.

\bibitem{chen2023q}
L.~Chen \emph{et~al.}, ``Q-DDCA: Decentralized dynamic congestion avoid routing in large-scale quantum networks,'' \emph{IEEE/ACM Transactions on Networking}, vol.~32, no.~1, pp.~368--381, 2024.

\bibitem{chen2024redp}
L.~Chen \emph{et~al.}, ``REDP: Reliable entanglement distribution protocol design for large-scale quantum networks,'' \emph{IEEE Journal on Selected Areas in Communications}, vol.~42, no.~7, pp.~1723--1737, 2024.

\bibitem{conti2024satellite}
A.~Conti, R.~Malaney, and M.~Z.~Win, ``Satellite-terrestrial quantum networks and the global quantum internet,'' \emph{IEEE Communications Magazine}, vol.~62, no.~10, pp.~34--39, 2024.

\bibitem{chiti2022mobile}
F.~Chiti, R.~Fantacci, R.~Picchi, and L.~Pierucci, ``Mobile control plane design for quantum satellite backbones,'' \emph{IEEE Network}, vol.~36, no.~1, pp.~91--97, 2022.

\bibitem{huang2023socially}
S.-M.~Huang, C.-Y.~Cheng, Y.-H.~Tsao, H.-C.~Wang, and J.-J.~Kuo, ``Socially-aware concurrent entanglement routing in satellite-assisted multi-domain quantum networks,'' in \emph{Proc. IEEE Int. Conf. Commun. (ICC)}, 2023, pp.~4924--4930.

\bibitem{gu2025quesat}
H.~Gu, R.~Yu, Z.~Li, X.~Wang, and G.~Xue, ``QuESat: Satellite-assisted quantum internet for global-scale entanglement distribution,'' in \emph{Proc. IEEE INFOCOM}, 2025, pp. 1--10.

\bibitem{williams2024scalable}
A.~Williams, N.~K.~Panigrahy, A.~McGregor, and D.~Towsley, ``Scalable scheduling policies for quantum satellite networks,'' in \emph{Proc. IEEE QCE}, vol.~1, 2024, pp. 1760--1769.

\bibitem{jennewein2025qeyssat}
T.~Jennewein \emph{et~al.}, ``QEYSSat 2.0---white paper on satellite-based quantum communication missions in Canada,'' \emph{Canadian Journal of Physics}, vol.~103, no.~4, pp.~328--376, 2025.

\bibitem{belenchia2022quantum}
A.~Belenchia \emph{et~al.}, ``Quantum physics in space,'' \emph{Physics Reports}, vol. 951, pp. 1--70, 2022.

\bibitem{werner1989quantum}
R.~F.~Werner, ``Quantum states with Einstein-Podolsky-Rosen correlations admitting a hidden-variable model,'' \emph{Phys. Rev. A}, vol. 40, no. 8, pp. 4277--4281, Oct. 1989.

\bibitem{xu2026smabsr}
Y.~Xu, X.~Liu, J.~Liu, T.~Taleb, Y.~Ji, and N.~Shiratori, ``SMAB-SR: A sleeping multi-armed bandit framework for secure routing in space--air--ground integrated networks,'' \emph{IEEE Transactions on Networking}, vol.~34, pp.~5066--5081, 2026.

\bibitem{puterman2014markov}
M.~L.~Puterman, \emph{Markov Decision Processes: Discrete Stochastic Dynamic Programming}. Hoboken, NJ, USA: John Wiley \& Sons, 2014.

\bibitem{weber1991addendum}
R.~R. Weber and G.~Weiss, ``Addendum to `On an index policy for restless bandits','' \emph{Advances in Applied Probability}, vol.~23, no.~2, pp.~429--430, 1991.

\bibitem{hsu2018age}
Y.-P. Hsu, ``Age of information: Whittle index for scheduling stochastic arrivals,'' in \emph{Proc. IEEE International Symposium on Information Theory (ISIT)}, 2018, pp.~2634--2638.

\bibitem{yates2021age}
R.~D.~Yates \emph{et~al.}, ``Age of information: An introduction and survey,'' \emph{IEEE Journal on Selected Areas in Communications}, vol. 39, no. 5, pp. 1183--1210, 2021.

\bibitem{shi2026quasar}
Y.~Shi, Z.~Wang, B.~Yuan, G.~Wu, and Z.~Zhao, ``QUASAR: Quantum Satellite Architecture and Routing Simulator,'' in \emph{Proc. 26th International Conference on Algorithms and Architectures for Parallel Processing (ICA3PP)}, 2026, to appear. [Online]. Available: https://arxiv.org/abs/2609.04920

\bibitem{chen2023simqn}
L.~Chen, K.~Xue, J.~Li, N.~Yu, R.~Li, Q.~Sun, and J.~Lu,
``SimQN: A network-layer simulator for the quantum network investigation,''
\emph{IEEE Network}, vol.~37, no.~5, pp.~182--189, 2023.

\bibitem{yehia2024connecting}
R.~Yehia, M.~Schiavon, V.~Marulanda Acosta, T.~Coopmans, I.~Kerenidis, D.~Elkouss, and E.~Diamanti, ``Connecting quantum cities: Simulation of a satellite-based quantum network,'' \emph{New Journal of Physics}, vol.~26, no.~7, p.~073015, 2024.

\bibitem{khatri2021spooky}
S.~Khatri \emph{et~al.}, ``Spooky action at a global distance: Analysis of space-based entanglement distribution for the quantum internet,'' \emph{npj Quantum Information}, vol.~7, no.~1, p.~4, 2021.

\bibitem{huang2024vacuum}
Y.~Huang \emph{et~al.}, ``Vacuum beam guide for large scale quantum networks,'' \emph{Physical Review Letters}, vol.~133, no.~2, p.~020801, 2024.

\bibitem{popa2024skyswapping}
A.-B.~Popa \emph{et~al.}, ``SkySwapping: Entanglement resupply by separating quantum swapping and photon exchange,'' \emph{Future Generation Computer Systems}, vol. 158, pp. 89--97, 2024.

\end{thebibliography}
\end{document}